\documentstyle[11pt,mrs2001,epsfig]{article}
\begin{document}
\title{Reconstructing cosmological fields using tessellation methods}

\author{W.E. Schaap, R. van de Weygaert}
\affil{Kapteyn Institute, PO Box 800, 9700 AV Groningen, The Netherlands.}

\begin{abstract}
Astronomical observations, physical experiments as well as computer
simulations often involve discrete data sets supposed to represent a
fair sample of an underlying smooth and continuous field.
Reconstructing the underlying fields from a set of irregularly sampled
data is therefore a recurring key issue in operations on astronomical
data sets. Conventional methods involve artificial filtering through a
grid or a smoothing kernel and fail to achieve an optimal result.
Here we describe a fully self-adaptive geometric method which does not
make use of artificial filtering, and which makes optimal use of the
available information.
\end{abstract}

\section{The Method}

Given a sample of field values at a discrete number of locations along
one dimension, various prescriptions are known for reconstructing the
field over the full spatial domain. In the left column of fig.~1 the
familiar 0th and 1st order approaches are shown.  In the right column
of fig.~1 the natural generalization to two dimensions is shown. A
natural choice for a 0th order multidimensional interpolation interval
is the Voronoi cell centered on any of the sample points \cite{sch00}.
Delimiting the influence region of a point, the Voronoi cell consists
of the region of space closer to the point than to any other point in
the sample. The 0th order interpolation scheme sets the value of the
field constant inside each Voronoi cell, its value everywhere in the
cell equal to that at its nucleus (top right fig.~1).  The 1st order
interpolation scheme is one in which the multi-dimensional linear
interpolation interval is the Delaunay cell (bottom right fig.~1).
Compelling evidence for the superb performance of this interpolation
scheme has been borne out by the work of Bernardeau \& van de Weygaert
\cite{ber96}. It yields a surprisingly accurate rendering of the
statistical distributions of a discretely sampled velocity field. This
was particularly significant as it uncovered systematic flaws in the
conventional grid-based interpolation schemes.

The implementation to any spatially discretely sampled field is rather
straightforward. However, an annoying complication occurs when seeking
to reconstruct the corresponding density field. The values of the
density field are not a priori specified or passed along with the
point sample. Instead, they have to be evaluated from the point
locations themselves. Schaap \& van de Weygaert \cite{sch00} showed
that in the case of 0th order interpolation it is indeed the inverse
of the volume of the Voronoi cell that represents the appropiate
density estimator.  On the other hand, the simple Voronoi cell volume
inverse does not suffice as density estimator in the Delaunay scheme:
the resulting integrated linearly interpolated density field does not
conserve mass.  The solution to this problem is to use the inverse of
the volume of the corresponding {\it contiguous} Voronoi cell, the
region defined by the union of all contiguous Delaunay tetrahedra
connected to a particular point.

\section{Application}

Cosmological N-body simulations provide an ideal template for
illustrating the virtues of our method. As may be appreciated from
fig.~2, they contain a large variety of objects, with diverse
morphologies and a large reach of densities, spanning over a vast
range of scales. The Voronoi-Delaunay method not only turns out to
provide a mere improvement in density estimation, but in particular
manages to deal fully self-adaptive -- i.e.~without any artificially
manipulated filtering parameters -- with two critically
important properties of cosmological density fields:\\
\hspace*{.6cm} $\bullet$ anisotropies (e.g.~filaments and walls)\\
\hspace*{.6cm} $\bullet$ hierarchies (from extended low density voids to compact, high-density clusters) \\
Conventional methods are usually tuned for uncovering a few aspects of
the full array of properties, and may therefore be insensitive to
unsuspected but intrinsically important structural elements.

The outstanding performance of our method is illustrated by fig.~2, in
which its performance (central panel) on a cosmological N-body
simulation is compared with that of a conventional grid-based
technique (right panel). The slice depicted contains two rich
clusters, some filamentary structures as well as almost empty
low-density regions.  Clearly the Delaunay method is able of
automatically resolving all the details present in the particle
distribution, while finer details are smoothed away by the grid-based
method: in the TSC reconstruction the high-density clusters appear to
be mere featureless blobs! In addition, low density regions are
rendered as slowly varying regions at moderately low values, while the
TSC reconstruction is plagued by annoying shot-noise effects. Also,
our method is able to reproduce sharp, edgy and clumpy filamentary
structures.

The presented example provides ample evidence of the promise of
tessellation methods for the aim of continuous field reconstruction.
We therefore feel that the application of tessellation methods to an
array of astronomical data analysis problems is more than warranted.
Indeed, it has prompted us to extend its operation to more complex,
noise- and selection-ridden situations.

\begin{figure} 
\plotone{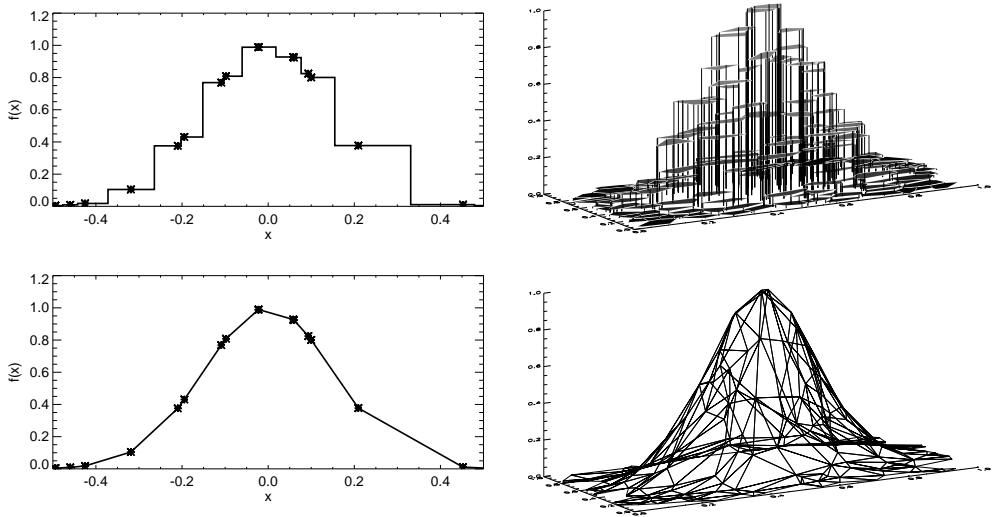}{13.14cm}
\caption{0th and 1st order interpolation in one (left column) and two (right column) dimensions.}
\end{figure}

\begin{figure} 
\vspace*{5cm}
\caption{Figure comparing the Delaunay density estimator (central panel) with a conventional grid-based TSC method (right panel) in analyzing a cosmological N-body simulation (left panel).}
\end{figure}

\vfill
\end{document}